\documentclass[prl,twocolumn,showpacs,preprintnumbers,amsmath,amssymb]{revtex4}

\usepackage{graphicx}
\usepackage{bm}
\usepackage{slashed}
\usepackage{bbm}
\usepackage{hyperref}

\newcommand{\Id}{\mathbbm{1}}

\newcommand{\indups}[1]{_{\mathrm{\scriptscriptstyle #1}}}

\newcommand{\Cusp}{\Gamma_{\indups{cusp}}}

\newcommand{\Op}{\mathcal{O}}

\begin{document}
\preprint{NORDITA-2011-94}
\preprint{UUITP-31/11}

\title{\mathversion{bold}Fine Structure of String Spectrum in $AdS_5\times S^5$}

\author{Konstantin Zarembo}
 \altaffiliation[Also at ]{{\it Department of Physics and Astronomy, Uppsala University, Sweden} and {\it ITEP, Moscow, Russia}.}
 \author{Stefan Zieme}%
\affiliation{%
Nordita,
Roslagstullsbacken 23, SE-106 91 Stockholm, Sweden
}%


\begin{abstract}
The spectrum of an infinite spinning string in $AdS_5$
does not precisely match the spectrum of dual gauge theory operators, interpolated to the strong coupling regime with the help of Bethe-ansatz equations. We show that the mismatch is due to interactions in the string $\sigma$-model which cannot be neglected even at asymptotically large 't~Hooft coupling.
\end{abstract}

\pacs{Valid PACS appear here}
\maketitle

According to the AdS/CFT correspondence 
\cite{Maldacena:1998re,Gubser:1998bc,Witten:1998qj},  $\mathcal{N}=4$ supersymmetric Yang-Mills theory (SYM) and string theory in $AdS_5\times S^5$ 
have a common spectrum that continuously interpolates between the loop-corrected dimensional analysis at weak coupling and the string oscillator spectrum at strong coupling. The complete integrability of the AdS/CFT system makes the non-perturbative interpolation amenable to an exact description by methods of Bethe ansatz  \cite{Beisert:2010jr}. The string interpretation of the spectrum, however, is quite subtle and our goal is to find a potential resolution of these subtleties.

We shall concentrate on a specific set of states related to twist-two operators $\mathop{\mathrm{tr}}\mathcal{Z}  D_+^S\mathcal{Z} $. Here $\mathcal{Z} $ is a complex scalar field in SYM and $D_+$ is the covariant derivative in light-cone direction.  The twist operators constitute presumably the most studied sector of the SYM spectrum \cite{Freyhult:2010kc}. Their anomalous dimensions scale logarithmically with spin: $\Delta _S(\lambda )-S\simeq 2\Cusp (\lambda )\ln S $, where the cusp anomalous dimension $\Cusp (\lambda )$ is a non-trivial function of the 't~Hooft coupling $\lambda =g_{\indups{YM}}^2N$, which can be computed non-perturbatively with the help of the Bethe-ansatz equations \cite{Beisert:2006ez}. 
On the string side, twist operators are described by a string spinning  in the Anti-de-Sitter space \cite{Gubser:2002tv}. When the spin is very large, the string becomes essentially infinite, extending all the way to the boundary. The energy density of this long string is equal to the cusp anomalous dimensions $\Cusp (\lambda )$.

We will be interested in the spectrum of small fluctuations on top of the long string, which are dual to operators with extra field insertions, schematically: $\mathop{\mathrm{tr}}\mathcal{Z} D_+^{l_1}\Psi_1 D_+^{l_2}\Psi _2\ldots D_+^{l_n}\mathcal{Z} $, where $\Psi _i$ can be a field strength, a fermion or a scalar. Each insertion corresponds to an elementary excitation above the ground state. The spectrum of elementary excitations can be found exactly \cite{Basso:2010in,Basso:2011rc} by solving the Bethe-ansatz equations \cite{Beisert:2006ez}, and should agree at strong coupling with the spectrum of the string in light-cone gauge. A detailed comparison reveals, however, several mismatches \cite{Giombi:2010bj}. Since the above operators have many uses, for instance they govern 
the collinear limits of scattering amplitudes \cite{Alday:2010ku}, it is important to  understand how these discrepancies are resolved.

The string oscillation modes in light-cone gauge are two-dimensional massive particles, whose interactions are suppressed by $1/\sqrt{\lambda }$. Let us list the $8_b+8_f$ modes of the string in $AdS_5\times S^5$, together with their masses and the corresponding worldsheet fields \cite{Frolov:2002av,Giombi:2009gd}:
\begin{eqnarray}
 {\rm AdS_3~transverse}~(\Phi):&&m^2=4
\nonumber \\
{\rm AdS_5~outside~AdS_3}~(X,X^*):&&m^2=2
\nonumber \\
{\rm S^5}~(Y^a,~a=1,\ldots ,5):&&m^2=0
\nonumber \\
{\rm Fermions}~(\Psi ^i,~i=1,\ldots ,4):&&m^2=1.
\end{eqnarray}
The fermions form four 2d Dirac spinors with eight degrees of freedom on shell.
The SYM spectrum, continued to strong coupling, consists of \cite{Basso:2010in}:
\begin{eqnarray}
 {\rm (Field~strength)}{}^\ell~(2):&&m^2=2\ell^2
\nonumber \\
{\rm Scalars}~(6):&&m^2=\tfrac{\sqrt{2}\lambda ^{1/4}}{\Gamma ^2(5/4)}\,\,{\rm e}\,^{-\sqrt{\lambda }/2}
\nonumber \\
{\rm Fermions}~(8):&&m^2=1,
\end{eqnarray}
where in brackets we indicated the number of degrees of freedom of each excitation. The agreement holds literally only for fermions. The heavy boson from $AdS_3$ is absent in the exact spectrum. The $S^5$ modes are massive, and there are six of them instead of five. A single insertion of the field strength ($\ell=1$) matches with the $AdS_5$ modes $X,X^*$. Multiple field strength insertions in addition can form bound states (Bethe strings) which are not directly visible in the string spectrum. The binding energy of the $\ell$-string,  
$E_\ell^{\rm bind}(p)\equiv \ell E_1({p}/{\ell})-E_\ell(p)$, is small at strong coupling \cite{Basso:2010in}:
\begin{equation}\label{bind}
 E_\ell^{\rm bind}(p)
 =\frac{\pi ^2\sqrt{2\ell^2+p^2}\left(\ell^2-1\right)\left(\ell^2+p^2\right)^4}{12\lambda  \ell^6\left(2\ell^2+p^2\right)}\,.
\end{equation}

These mismatches indicate that the string spectrum is strongly affected by interactions,  even at asymptotically large $\lambda $. The mechanism for mass generation of the $S^5$ modes, which is clearly non-perturbative, was understood in \cite{Alday:2007mf} and later confirmed by Bethe-ansatz calculations \cite{Basso:2008tx}. Since all other modes are massive, the low-energy effective theory at worldsheet momenta $p\ll 1$ is the $O(6)$  $\sigma$-model, which is asymptotically free and generates a mass gap through dimensional transmutation. 
The first correction to the dispersion relation of the $S^5$ modes that goes beyond the low-energy approximation was computed recently both in string theory \cite{Giombi:2010bj} and from the Bethe-ansatz equations \cite{Basso:2010in}. In both cases the result has the form:
\begin{equation} \label{defc}
 \epsilon^2 (p)=p^2- {c}\lambda^{-1/2} \ p^4+\Op (\lambda^{-1}) \,,\quad (p\gg m),
\end{equation}
but the coefficient is different:
 $c_{\rm string}=7/3\approx 2.333$ \cite{Giombi:2010bj} while  $c_{\indups {BA}}=\sqrt{2}\Gamma ^4(1/4)/(24\cdot 3^{1/4}\pi )\approx 2.463$  \cite{Basso:2010in}. This is quite surprising since the $p^4$ term is relevant in the UV at $p^2\gg m^2$ when perturbation theory is supposed to work.

We thus shall address the following questions: $(i)$ Can the light $AdS$ scalars form bound states? $(ii)$ What is the fate of the heavy $AdS$ mode? and $(iii)$ What accounts for the difference between perturbative and exact values of the
$p^4$ coefficient in the dispersion relation of the $S^5$ modes? Our starting point is the string $\sigma$-model in light-cone gauge, with the action expanded to quartic order in the fields \cite{Giombi:2009gd,Giombi:2010bj}:
\begin{widetext}
\begin{eqnarray} \label{Lagrange}
\mathcal{L}&=&(\partial_\alpha \Phi)^2 +4\Phi^2 + |\partial_\alpha X|^2  +2|X|^2+ (\partial_\alpha Y^a)^2 +  2i\bar{\Psi}(\slashed{\partial}+\Id)\Psi
\nonumber \\
&&
 + 2 \Phi \left[ (\partial_t \Phi)^2 - (\partial_s \Phi)^2+ (\partial_t Y^a)^2 - (\partial_s Y^a)^2
 -2 |\nabla_s X  |^2 
+ 2i\nabla_s \bar{\Psi}  \Pi_+\Psi+2i\bar{\Psi}\Pi_{-} \nabla_s  \Psi  \right] \nonumber\\
 &&+2 i Y^a \left[ \nabla_s \bar{\Psi}  \Pi_+\rho ^{a6}\Psi
 			-\bar{\Psi} \Pi_{-} \rho ^{a6}\nabla_s  \Psi 	
		 \right] 
+ 2i \partial_t Y^a\bar\Psi \gamma_0 \Pi_+\rho^{a6} \Psi
+ 2 \nabla_s X \Psi^\mathrm{t} \Pi_+\rho^6 \Psi
- 2  \nabla_s X^* \bar\Psi \Pi_- \rho^\dagger_6\bar\Psi^\mathrm{t} \nonumber \\
&&+ 2 \Phi^2 \left[ (\partial_\alpha \Phi)^2 + \tfrac{2}{3} \Phi^2
+(\partial_\alpha Y^a)^2 
+ 4 |\nabla_s X |^2  \right] 
- \tfrac{1}{2}(Y^a)^2(\partial_\alpha Y^a)^2 
\nonumber \\
&&
+i\left[(Y^a)^2-4\Phi^2\right](\nabla_s \bar{\Psi}  \Pi_+\Psi+\bar{\Psi} \Pi_{-} \nabla_s  \Psi ) 
+4 i \Phi Y^a
(\bar{\Psi}\Pi_{-}\rho ^{a6} \nabla_s  \Psi - \nabla_s \bar{\Psi}  \Pi_+\rho ^{a6}\Psi)
 \nonumber\\
&&+6\Phi (\nabla_s X^*\bar\Psi\Pi_- \rho^\dagger_6\bar\Psi^\mathrm{t} - \nabla_s X \Psi^\mathrm{t} \Pi_+ \rho^6\Psi )
 - 2i Y^a \partial_t Y^b  \bar\Psi \gamma_0 \Pi_+\rho^{ab} \Psi 
+(\bar\Psi\gamma_0 \Pi_+ \rho ^{a6}\Psi)^2
- (\bar\Psi_i \gamma_0 \Pi_+ \Psi)^2.	 
\end{eqnarray}
\end{widetext}
The notations follow \cite{Giombi:2009gd,Giombi:2010bj}, with the exception of fermions which we have brought into the 2d Dirac form. We use $\gamma^\alpha =(-\sigma ^1,\sigma ^3) $ as the basis of 2d Dirac matrices, $\Pi _\pm=(1\pm\gamma ^s)/2$, 
 $\nabla_s=\partial _s-1$. $\rho^a ,\rho ^6$ are the $4\times 4$ chiral components of the 6d Dirac matrices, and $\rho ^{AB}=\rho ^{[A\,\dagger }\rho ^{B]}$. The action is written in Euclidean signature and is multiplied by an overall factor of $\sqrt{\lambda }/4\pi$.
 
We begin with the bound states of the light scalars $X$. The small binding energy makes these bound states non-relativistic at strong coupling. The binding energy can thus be derived from a low-energy effective Hamiltonian. The effective potential is determined by  matching the $XX\rightarrow XX$ scattering amplitude in the $\sigma $-model to the  Born amplitude
$\mathcal{M}(q)=-2\left(2\sqrt{2}\right)^2\int_{}^{}dx\,\,{\rm e}\,^{-iqx}V_{\rm eff}(x)$, where $q$ is the momentum transfer in the $t$-channel. The $XX$ scattering at tree level proceeds through the exchange of the $\Phi$ boson. The $t$- and $u$-channel diagrams combine to 
\begin{equation}
 \mathcal{M}=\frac{16\pi}{\sqrt{\lambda}}+\Op\left(q^2\right)~ \Longrightarrow~ V_{\rm eff}(x)=-\frac{\pi }{\sqrt{\lambda }}\,\delta (x).
\end{equation}
The effective potential is thus an attractive delta function, which has one discrete level with energy $\pi ^2/2\sqrt{2}\lambda $.
The solution of the Schr\"odinger equation for $\ell$ particles interacting pairwise via the delta-function potential is also known  \cite{McGuire:1964zt}. There is a single bound state for each $\ell$ with binding energy
\begin{equation}
 E_\ell^{\rm bind}=\frac{\pi ^2\sqrt{2}\ell\left(\ell^2-1\right)}{24\lambda }\,,
\end{equation}
that agrees nicely with (\ref{bind}) in the static regime.

To address the fate of the heavy bosonic mode we need to study quantum corrections to its propagator. Since the heavy boson is a factor of two heavier than the fermions, it may dissolve in the continuum of the two-particle states, as it happens in a similar context in $AdS_4$, \cite{Zarembo:2009au}. The pole of the boson propagator disappears by moving onto the unphysical sheet of the complex energy plane. Consider first one-loop corrections:
\begin{equation}\label{G-1}
 G^{-1}(p)=p^2+4+\frac{C}{\sqrt{\lambda }}\,\sqrt{p^2+4}+\ldots \,,
\end{equation}
where we assume that the momentum is very close to the threshold: $p^2\rightarrow -4$ and keep only the most singular part of the polarization operator. The would-be pole lies at the edge of the two-particle continuum and may disappear,  depending on the sign of $C$. If $C$ is negative,
the two terms in (\ref{G-1})  have opposite signs for $0>p^2 >-4$. They cancel at some $p^2$ and
the pole remains on the physical sheet just below the threshold.  However, the explicit calculation \cite{Giombi:2010bj} shows that $C$ is positive. As we shall see, the positivity of $C$ is a simple consequence of unitarity. The pole then disappears (it moves into the unphysical sheet), and the only remaining singularity of the Green's function is the two-particle cut. If (\ref{G-1}) were exact, we would conclude that the heavy boson does not exist as an independent excitation.
But there are other corrections to the boson propagator that should also be taken into account. First, the mass-shell conditions for the boson and fermions get loop corrections \cite{Giombi:2010bj}. Unless $\epsilon _b(p)=2\epsilon _f(p/2)$, these corrections shift the pole away from the threshold, and then (\ref{G-1}) will have an isolated zero. The one-loop correction to the boson dispersion relation is the constant part of the polarization operator, while  
the compensating correction to the fermion mass-shell condition, $p^2=1+\epsilon _{f(1)}^2/\sqrt{\lambda }$, affects the polarization operator at two loops:
\begin{equation}
 \mathop{\mathrm{Disc}}\Pi_{(2)} {\rm }(p)=C_{(2)}
 \sqrt{p^2+4}+\frac{D_{(2)}}{\sqrt{p^2+4}}\,,
\end{equation}
where $D_{(2)}$ must be equal to $2\epsilon ^2_{f(1)}C_{(1)}$, in order for the second term to combine into the perturbative expansion of the square root in (\ref{G-1}) . But what if $D_{(2)}\neq 2\epsilon ^2_{f(1)}C_{(1)}$? And also why $D_{(1)}=0$? If $D\neq 0$, the heavy particle would decay rather than dissolve in the continuum.  The unexpected softening  of the threshold
singularity, in fact, 
can be traced to the structure of the boson-fermion vertex \cite{Giombi:2010bj}: the threshold singularity of the polarization operator is related by unitarity to the $\Phi \rightarrow \bar{\Psi }\Psi $ amplitude, fig.~\ref{htoff}a: 
\begin{figure}
 \centerline{\includegraphics[width=7cm]{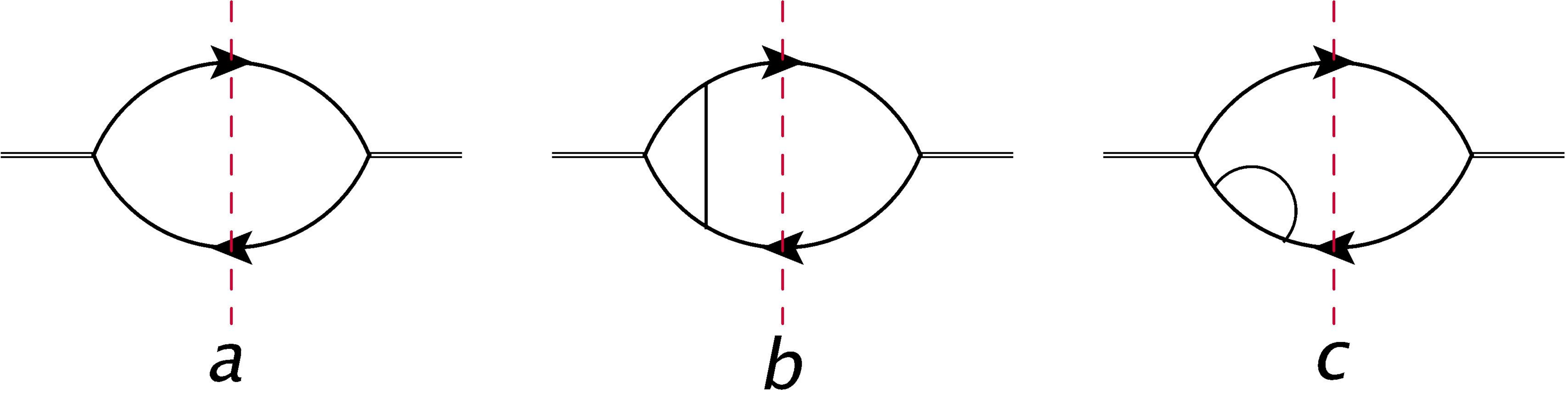}}
\caption{\label{htoff} Possible unitarity cuts of the fermion loop in the polarization operator of the heavy boson.}
\end{figure}
\begin{equation}
 \mathop{\mathrm{Im}}\Pi (p)=-\frac{\left|\mathcal{A}_{h\rightarrow ff}(p)\right|^2}{4p^2\sqrt{p^2+4}}\,,
\end{equation}
where $\mathcal{A}_{h\rightarrow ff}(p)$ is the $\Phi \bar{\Psi }\Psi $ vertex with both fermions on-shell: 
$\mathcal{A}_{h\rightarrow ff}(p)=\bar{v}(k)\Gamma(p;k,p-k)u(p-k) $. The fermion and antifermion wavefunctions
\begin{equation}
 u(k)=\frac{1}{\sqrt{k_t}}\begin{pmatrix}
   k_t \\ 
    {k_s-i} \\ 
 \end{pmatrix},\,\,\,
 \bar{v}(k)=\frac{1}{\sqrt{k_t}}\begin{pmatrix}
   -k_s+i, & {k_t} 
 \end{pmatrix},
\end{equation}
are such that $\bar{v}(k_t,-k_s)u(k_t,k_s)=2k_s$. At tree level 
$\Gamma (p;k,p-k)=4k_s\gamma ^s+4p_s\Pi _--4i$, from which it immediately follows that in the rest frame of the heavy boson ($p_s=0$) the amplitude vanishes at threshold. Consequently, $\mathop{\mathrm{Im}}\Pi (p)\sim ({p^2+4})^{1/2}$. At two loops there are two possible two-particle cuts, fig.~\ref{htoff}b and c. The cut \ref{htoff}c produces the requisite coefficient $D_{(2)}$. The cut \ref{htoff}b should only contain the square-root singularity. For this to happen, the amplitude $\mathcal{A}_{h\rightarrow ff}(p)$ must vanish on-shell at the one-loop order. 
Extrapolating this reasoning to higher orders we  conclude that the heavy boson disappears from the spectrum if two conditions are satisfied:
\begin{equation}
 \epsilon _b(p)=2\epsilon _f(p/2),\qquad \bar{v}\Gamma (p;k,p-k)u=0 \,.
\end{equation}
where $\epsilon _b(p)$ and $\epsilon _f(p)$ are exact on-shell energies of the heavy boson and of the fermions.
The first condition is satisfied at one loop \cite{Giombi:2010bj}. We have checked that the second condition is also fulfilled by calculating all one-loop corrections to the vertex function in the rest frame of the heavy boson, where $p_s=0=k_s$. The Feynman diagrams that contribute have three different topologies 
and give in total sixteen diagrams if one uses vertices from the Lagrangian (\ref{Lagrange}). 
Full details of the calculation will be presented in 
\cite{ZZfuture}.
Eventually, all one-loop vertex corrections are proportional to the unit matrix in the Dirac indices or vanish in the threshold kinematics. This happens diagram-by-diagram even before momentum integration, but exploits symmetry properties of the integrands. As a consequence, the decay amplitude $\bar{v}\Gamma(p;k,p-k)u $
vanishes on-shell because $\bar{v}(k)u(k)=0$.

We now turn to the scalar modes on $S^5$. These modes have no mass term in the Lagrangian since they are Goldstone bosons of the $SO(6)$ symmetry, broken to $SO(5)$ by a choice of the reference point on $S^5$. In two dimensions Goldstone bosons cannot exist \cite{Coleman:1973ci,Mermin:1966fe}, which in perturbation theory is reflected in IR divergences. The theorems of \cite{Elitzur:1978ww,David:1980rr} suggest that the IR divergences cancel in $SO(6)$-invariant quantities. For instance the free energy $\Gamma _{\indups {cusp}}(\lambda )$  is IR finite at least to the two-loop order  \cite{Roiban:2007dq}. In a non-invariant observable,  IR logarithms will start to appear at a certain order of perturbation theory, which we denote by $n$. The IR logarithms are cut off at the scale of mass generation in the $O(6)$ $\sigma$-model, and thus the $n$-loop IR divergence is proportional to $\lambda ^{-n/2}\ln(\Lambda /m)\sim \lambda ^{-(n-1)/2}$, which is of the same order as the IR-free contribution at $(n-1)$ loops. Multiple logarithms at higher orders are all $\Op(\lambda ^{-(n-1)/2})$ thus invalidating perturbation theory beyond $(n-2)$ loops.  Although the constant $c$ defined in (\ref{defc}) is IR  finite at one loop \cite{Giombi:2010bj}, it cannot be  reliably computed in perturbation theory because of two-loop IR divergences. However, at large $\lambda $ the UV and IR scales are highly separated, and one can consistently calculate an effective action for the IR modes by integrating out the heavy  $\sigma $-model fields. In this interpretation $c_{\rm string}$ should be viewed as a coefficient of a dimension-4 operator in the low-energy effective  action at the matching scale $\mu \sim 1$, rather than a constant in the physical dispersion relation. The RG evolution down to scales $\mu \sim m$ should account for the difference. The effect of the RG evolution, however, is numerically small (only $6\%$). We explain this by the accuracy of the large-$N$ expansion for the $O(N)$ $\sigma$-model at $N=6$. At infinite $N$, the mean-field theory is exact and the dispersion relation is read off from the action.  We will compute the next order in $1/N$ to see if this improves the agreement.

At the renormalizable level, the effective action for the  $S^5$ string modes is  the $O(6)$ $\sigma$-model \cite{Alday:2007mf}:
\begin{equation}\label{l2}
 \mathcal{L}^{\rm eff}_2=\frac{\sqrt{\lambda}}{4\pi} \,\partial \mathbf{n}\cdot \partial \mathbf{n}.
\end{equation}
In the full string action the 2d Lorentz invariance is broken by interactions. We thus expect that integrating out heavy worldsheet fields will induce Lorentz-symmetry breaking terms in the effective action for the $O(6)$ modes, starting with operators of dimension four. There are two possible structures:  $(\partial\mathbf{n} \,\partial\mathbf{n})^2$ and $\partial^2\mathbf{n} \,\partial^2\mathbf{n}$. An operator of the first type is generated at tree level from integrating out the $\Phi $ field in (\ref{Lagrange}). The second structure arises at one loop and  can be extracted from the polarization operator of the Goldstone modes computed in \cite{Giombi:2010bj}. One should be careful to avoid double-counting since the Goldstone modes  contribute to the polarization operator through self-coupling. Subtracting the self-interaction \cite{ZZfuture}, we obtain:
\begin{eqnarray}\label{l4}
\mathcal{L}^{\rm eff}_4&=&  -\frac{\sqrt{\lambda }}{16\pi }\left(\partial \mathbf{n}\circ\partial \mathbf{n}\right)^2
-\frac{1}{6\pi }\,\left(\partial\cdot \partial \mathbf{n}\right)^2
\nonumber \\
&&
 -\frac{1}{4\pi }\,\partial \cdot \partial \mathbf{n}\,\partial \circ\partial \mathbf{n}
 -\frac{1}{12\pi }\,\left(\partial\circ \partial \mathbf{n}\right)^2,
\end{eqnarray}
where we have introduced the notation $u\circ v =u_0v_0-u_1v_1$ to parameterize the violation of  the rotational symmetry and used $u \cdot v=u_0v_0+u_1v_1$ for the usual scalar product. 

The leading order of the large-$N$ expansion for the effective Lagrangian (\ref{l2}), (\ref{l4}) is equivalent to the random phase approximation.  For the correct large-$N$ power counting  $\sqrt{\lambda }$ should scale linearly with $N$, and it is convenient to introduce an effective coupling $\kappa =2\pi N/\sqrt{\lambda }$, which stays finite in the large-$N$ limit. Since we are ultimately interested in the leading Lorentz-violating term in the dispersion relation (\ref{defc}),  we can set $p_0^2+p_1^2$ to zero in the dimension-4 part of the action and replace $p_0^2-p_1^2$ by $-2p_1^2$. After imposing the condition $\mathbf{n}^2=1$ by a Lagrange multiplier and applying the Hubbard-Stratonovich transformation to disentangle the quartic terms, we obtain:
\begin{eqnarray}
  \mathcal{L}_{O(6)}&=&\frac{N}{2\kappa }
 \left[
 \left(\partial \mathbf{n}\right)^2+\left(m^2+i\sigma \right)\mathbf{n}^2
 -i\sigma -\frac{1}{M_1^2}\left(\partial _1^2\mathbf{n}\right)^2
 \right.
\nonumber \\
&&\left. \vphantom{\frac{1}{M_1^2}}
 -\phi \left(\partial _1\mathbf{n}\right)^2+\kappa M_2^2\phi ^2
 \right],
\end{eqnarray}
with $M^2_1=3\sqrt{\lambda }/4$ and $M_2^2=\sqrt{\lambda }/48\pi $. Integrating out $\mathbf{n}$ we get an effective action for $\sigma $ and $\phi $, whose minimization in $\sigma $ yields a gap equation that determines the dynamically generated mass $m$.  The $1/N$ correction to the $\mathbf{n}\mathbf{n}$ propagator then determines the coefficient $c$ in the dispersion relation (\ref{defc}). The details of the rather lengthy calculation will be presented in a future publication \cite{ZZfuture}. The final result, however, is quite simple:
\begin{equation}
 \frac{c_{\indups{RPA}}}{\sqrt{\lambda }}=\left(1-\frac{6\ln 2-3\ln 3}{N}\right)\frac{1}{M_1^2}+\frac{\ln 2}{4\pi N}\,\,\frac{1}{M_2^2}\,.
\end{equation}
Potential UV divergences cancel in $c$, which means that all the contributions come from modes with  $p\sim m$. Substituting numbers we get $c_{\indups{RPA}}=2(2+\ln 6)/3\approx 2.528$, which improves the mean-field prediction $2.333$ by $50\%$. To compute the exact value $c_{\rm exact}=2.463$ one has to solve the model non-perturbatively. 
 
In conclusion, we managed to reconcile mismatches between the string $\sigma$-model and the Bethe ansatz solution, at least at the qualitative level. The origin of these mismatches can be traced to the non-perturbative nature of interactions on the string worldsheet. Even if the $\sigma$-model coupling, $2\pi /\sqrt{\lambda }$, is very small, interactions cannot be neglected and lead to rearrangements in the perturbative string spectrum. Since the string $\sigma$-model in $AdS_5\times S^5$ is integrable, the spectral mutations are captured by the Bethe ansatz equations in this particular case, but we believe that the phenomena studied in this paper are more general and are not specific to the $AdS_5\times S^5$ background. 
\begin{acknowledgments}
We would like to thank B.~Basso, S.~Giombi, N.~Gromov, J.~Maldacena, S.~Nowling, R.~Ricci, R.~Roiban, A.~Tseytlin and P.~Vieira for illuminating discussions and for comments on the draft. 
 The work of K.Z.~was supported in part by the Swedish Research Council under contract 621-2007-4177, in part by the ANF-a grant 09-02-91005, in part by the RFFI grant 10-02-01315, and in part by the Ministry of Education and Science of the Russian Federation under contract 14.740.11.0347.
\end{acknowledgments}

\end{document}